# Bifurcation of the Kirkendall marker plane and the role of Ni and other impurities on the growth of Kirkendall voids in the Cu–Sn system

Varun A Baheti, Sanjay Kashyap, Praveen Kumar, Kamanio Chattopadhyay and Aloke Paul[1*]
Department of Materials Engineering, Indian Institute of Science, Bangalore 560012
*Corresponding author: E–mail: aloke@materials.iisc.ernet.in,
Tel.: 918022933242, Fax: 91802360 0472

**Abstract**

The presence of bifurcation of the Kirkendall marker plane, a very special phenomenon discovered recently, is found in a technologically important Cu–Sn system. It was predicted based on estimated diffusion coefficients; however, could not be detected following the conventional inert marker experiments. As reported in this study, we could detect the locations of these planes based on the microstructural features examined in SEM and TEM. This strengthens the concept of the physico–chemical approach that relates microstructural evolution with the diffusion rates of components and imparts finer understanding of the growth mechanism of phases. The estimated diffusion coefficients at the Kirkendall marker planes indicates that the reason for the growth of the Kirkendall voids is the non–consumption of excess vacancies which are generated due to unequal diffusion rate of components. Systematic experiments using different purity of Cu in this study indicates the importance of the presence of impurities on the growth of voids, which increases drastically for $\geq$ 0.1 wt.% impurity. The growth of voids increases drastically for electroplated Cu, commercially pure Cu and Cu(0.5 at.%Ni) indicating the adverse role of both inorganic and organic impurities. Void size and number distribution analysis indicates the nucleation of new voids along with the growth of existing voids with the increase in annealing time. The newly found location of the Kirkendall marker plane in the $Cu_3Sn$ phase indicates that voids grow on both the sides of this plane which was not considered earlier for developing theoretical models.

**Keywords:** Interdiffusion; Diffusion mechanism; Microstructure; The Kirkendall effect.

1. Introduction

In the last decade, the discovery of splitting of the Kirkendall markers into more than one plane (bifurcation and trifurcation) led to the development of many new theories in the solid state diffusion [1]. One of such theories, a physico–chemical approach [2] (developed by one the authors of this manuscript) relates the microstructural evolution with the rates of diffusing components and helps to predict the location of the Kirkendall marker planes in a particular system. This was applied to the technologically important Cu–Sn system [3] since the growth of the brittle $Cu_3Sn$ and $Cu_6Sn_5$ intermetallic phases along with the Kirkendall voids (by Frenkel effect) in the interdiffusion zone of Cu under bump metallization (UBM) and Sn–based solder is a major concern for the electronic industries. Analysis of the microstructural evolution with respect to the location of the Kirkendall marker plane is an important aspect for understanding the physico–mechanical properties of these brittle phases and thereby the reliability of an electronic component. Numerous articles are published every year in the Cu–Sn or Cu–Solder (Sn–based) systems studying these aspects and the details of these can be found in References [3-8] and references therein. However, a clear understanding of the growth mechanism of these brittle phases is yet to be developed. Interestingly, Paul et al. [3] reported a mismatch between the predicted (based on physico–chemical approach [2]) and experimentally found locations of the Kirkendall marker plane. Identifying the locations of these planes is important for understanding the finer details of the diffusion–controlled growth process of both the phases. A single Kirkendall marker plane is found (inside the $Cu_6Sn_5$ phase [3-5]) in these previously studied Cu/Sn diffusion couples. However, based on the estimated diffusion coefficients from incremental diffusion couples, Paul et al. [3] predicted a bifurcation of the Kirkendall marker plane *i.e.* the presence of the Kirkendall marker plane in both the phases viz. $Cu_3Sn$ and $Cu_6Sn_5$. Latter the same was validated following a different theoretical analysis by Svoboda et al. [9].

At the same time, numerous articles are published on the growth of Kirkendall voids as reported in References [6-8, 10-14] and references therein. Due to drive for miniaturization, the growth of voids in the $Cu_3Sn$ phase is one of the main reasons of electro–mechanical failure in micro–electronic components. Immediately after the discovery of the Kirkendall effect [15], researchers could correlate the growth of voids in an interdiffusion zone with this effect [10-14]. A flux of vacancies is created because of a difference in the diffusion rates of components, which are supersaturated and nucleated heterogeneously if not absorbed by the sinks such as dislocations, grain boundaries and interfaces [11, 13, 14]. The presence of impurities is known to play an adverse role on the growth of these voids [6, 11]. Because of industrial relevance, most of the studies are conducted with electroplated Cu in which impurities can be included from the electroplating bath leading to the very high growth rate of these voids in the interdiffusion zone of Cu UBM and Sn–based solder [6-8]. However, confusion exists over the exact role of impurities since these studies are not compared extensively for the known and different concentration of impurities in Cu.

Therefore, the aim of the present manuscript is two–fold. The first aim is to examine if a bifurcation of the Kirkendall marker plane is indeed present (as predicted [3]) in the Cu–Sn system. If it is present, then what is the reason for not detecting it [3-5] in the previous experiments? Following, based on the estimation of the diffusion coefficients at the Kirkendall marker plane(s); the second aim is to understand the role of impurities on the growth of Kirkendall voids by considering Cu with different concentration of impurities. This is analyzed based on the effect of concentration of impurities on void size distribution.

## 2. Results and Discussion

The experiments are conducted following the diffusion couple technique. The details of making diffusion couples can be found in Refs. [1, 16]. Other important details are incorporated during discussion of the results.

### 2.1 Bifurcation of the Kirkendall marker plane in the Cu–Sn system

It should be noted here that the presence of more than one Kirkendall marker plane is a special phenomenon and found only in very few systems. It allows developing a finer understanding of the phenomenological diffusion process [1]. There are different ways to detect the location of this plane in an interdiffusion zone. The conventional method is of course by the use of inert particles [1]. On the other hand, as established based on the physico–chemical approach [2], the microstructural features efficiently indicates these locations without using any inert particles. Since the experiments using the inert particles failed to show the presence of bifurcation of the Kirkendall marker plane in this system [3-5] (if predicted correctly [3]), the microstructural evolution of both the phases viz. $Cu_3Sn$ and $Cu_6Sn_5$ is examined for our analysis.

The SEM micrographs of Cu/Sn diffusion couple are shown in Figure 1 revealing the microstructure in the $Cu_3Sn$ and $Cu_6Sn_5$ phases. Before going for further explanation, it is necessary to understand the growth mechanism of the phases following the physico–chemical approach [2] and what kind of microstructural evolution is expected depending on the location of the Kirkendall marker plane. This can be explained with the help of a schematic diagram as shown in Figure 2a.

The reaction–dissociation of the components at different interfaces and therefore the morphological evolution is described below qualitatively. The quantitative mathematical analysis of the diffusion coefficient dependent on the microstructural evolution can be learnt from Ref. [2].

*The growth of the $Cu_3Sn$ phase:*

(i)  Cu is released from Cu end–member at the interface I (Cu/$Cu_3Sn$). It diffuses through $Cu_3Sn$ and then reacts with $Cu_6Sn_5$ at the interface II ($Cu_3Sn$/$Cu_6Sn_5$) for the growth of $Cu_3Sn$ from the same interface.

(ii)  Sn dissociates from $Cu_6Sn_5$ at the interface II to produce $Cu_3Sn$ and Sn. The same Sn then diffuse through $Cu_3Sn$ and reacts with Cu at the interface I to produce $Cu_3Sn$.

*The growth of the $Cu_6Sn_5$ phase:*

(iii)  $Cu_3Sn$ dissociates at the interface II to produce $Cu_6Sn_5$ and Cu. The same Cu then diffuse through $Cu_6Sn_5$ and reacts with Sn at the interface III ($Cu_6Sn_5$/Sn) to produce $Cu_6Sn_5$.
(iv)  Sn is released from the Sn end–member at the interface III. It diffuses through $Cu_6Sn_5$ and then reacts with $Cu_3Sn$ at the interface II for the growth of $Cu_6Sn_5$.

Note that the reaction–dissociation process controls the microstructural evolution; however, the growth of phases depends on the diffusion rates of components. From the discussion above, it must be evident that (depending on the diffusion rates of Cu and Sn) the $Cu_3Sn$ phase grows from interfaces I and II, while the $Cu_6Sn_5$ phase grows from interfaces II and III. $Cu_3Sn$ at the interface I (with sublayer thickness of $\Delta x^{I}_{Cu_3Sn}$) and $Cu_6Sn_5$ at the interface III (with sublayer thickness of $\Delta x^{III}_{Cu_6Sn_5}$) grow without getting consumed by the neighbouring phases. However, the growth process is complicated at the interface II as both the phases try to grow at the cost of the other phase. The presence of a single or bifurcation of the Kirkendall marker plane depends on the growth and consumption rates of the phases at the interface II. If both the phases have their growth rate higher than the consumption rate, then $\Delta x^{II}_{Cu_3Sn}$ and $\Delta x^{II}_{Cu_6Sn_5}$ will have positive values. In such a situation, both the phases will grow with two sublayers. Therefore, the bifurcation of the Kirkendall marker ($K_1$ in $Cu_3Sn$ and $K_2$ in $Cu_6Sn_5$) and hence splitting of inert marker particles should be found in this Cu–Sn system [2]. Since both the phase layers grow differently from two different interfaces, a duplex morphology should be found demarcated by $K_1$ and $K_2$, as shown in schematic Figure 2a. On the other hand, if the thickness of any of the sublayers $\Delta x^{II}_{Cu_3Sn}$ or $\Delta x^{II}_{Cu_6Sn_5}$ is negative (*i.e.* consumption rate of a phase at the interface II is higher than the growth rate), a single Kirkendall marker plane should be found. For example, if $\Delta x^{II}_{Cu_6Sn_5}$ is positive and $\Delta x^{II}_{Cu_3Sn}$ is negative, then the marker plane will be present only in the $Cu_6Sn_5$ phase. A negative value of $\Delta x^{II}_{Cu_3Sn}$ means that the sublayer $\Delta x^{II}_{Cu_3Sn}$ along with some part of $Cu_3Sn$ which is grown from interface I ($\Delta x^{I}_{Cu_3Sn}$) is consumed because of the growth of $Cu_6Sn_5$ at the interface II. In such a situation, a duplex morphology in the $Cu_6Sn_5$ phase (demarcated by $K_2$) and one type of morphology (because of growth only from interface I) is expected to be found in the $Cu_3Sn$ phase. This will lead to the presence of a single Kirkendall marker plane in this system.

As already mentioned, the splitting of the Kirkendall marker plane is a special phenomenon and found only in very few systems. Most of the systems, in general, grow with a single Kirkendall marker plane. Therefore, as a general trend, efforts are not made to locate more than one marker plane. On the other hand, if the diffusion parameters in different phases are known, one can calculate the location of the Kirkendall marker planes in a particular system [2]. Paul et al. [3] estimated these parameters in $Cu_3Sn$ and $Cu_6Sn_5$ by incremental diffusion couple experiments in which couples were prepared such that a single phase grows in the interdiffusion zone. For example, Cu and $Cu_6Sn_5$ were coupled for the growth of $Cu_3Sn$, whereas $Cu_3Sn$ and Sn were coupled for the growth of $Cu_6Sn_5$ [3]. Subsequently, these diffusion parameters were used to calculate the thickness of the sublayers in both the phases viz. $Cu_3Sn$ and $Cu_6Sn_5$ in a Cu/Sn diffusion couple. Surprisingly, positive values of all the sublayers were calculated indicating the presence of a bifurcation of the Kirkendall marker plane in the Cu–Sn system [3]. However, the inert markers could locate this plane only in the $Cu_6Sn_5$ phase showing the presence of a single Kirkendall marker plane [3-5]. To examine this disparity, as an alternate method, microstructure evolution is examined in this study to locate the Kirkendall marker plane.

Figure 1a is the BSE micrograph of the interdiffusion zone of Cu/Sn diffusion couple which shows the presence of both the phases $Cu_3Sn$ and $Cu_6Sn_5$ after annealing at 200 °C for 81 hrs. The grains of the $Cu_6Sn_5$ phase could be resolved by increasing the contrast–brightness of the same image (in which the $Cu_3Sn$ phase is not visible). A duplex morphology is clearly visible inside the $Cu_6Sn_5$ phase demarcated by the Kirkendall marker plane, as shown by the dashed line. Previously, a similar location of the marker plane was detected by the use of inert markers [3, 16]. A polarized light optical micrograph of a Cu/Sn diffusion couple annealed at 215 °C for 1600 hrs, as shown in Figure 1b (kindly provided by Dr. A.A. Kodentsov, Eindhoven University of Technology, The Netherlands), also show a similar nature of the grain morphology inside the $Cu_6Sn_5$ phase in which Kirkendall markers (*i.e.* inert particles) were also found along the dashed line [16]. This indicates that $\Delta x^{II}_{Cu_6Sn_5}$ could grow despite being consumed by $Cu_3Sn$ at the interface II (Figure 2a). Since $Cu_6Sn_5$ grains cover from the interfaces (II and III) to the Kirkendall marker plane (K$_2$), it is evident that once nucleated grains grew continuously without further nucleation. This is common in majority of the systems with intermetallic compounds, most probably because of the high activation energy barrier for nucleation [1]. The grain morphology of the Cu$_3$Sn phase could not be resolved in the image shown in Figure 1a. However, the polarized light micrograph, as shown in Figure 1b, indicates that the elongated grains might be present covering almost the whole Cu$_3$Sn phase layer. The grains in the same phase could be faintly detected in a BSE image of Cu/Sn diffusion couple annealed at 200 °C for 400 hrs, as shown in Figure 1c (grain boundaries are marked by dashed lines in a focused SE image). It is difficult to resolve the morphology at very near to the Cu/Cu$_3$Sn interface (interface I); however, rest of the Cu$_3$Sn phase thickness is covered by long grains covering almost the whole phase layer. The Kirkendall plane (denoted by K$_1$ in schematic Figure 2a) should not be found in the Cu$_3$Sn phase if the elongated grains indeed cover the whole Cu$_3$Sn phase layer. Since inert particles used to detect the location of the Kirkendall marker plane were not found in this phase, it was accepted that only one Kirkendall marker plane (inside the Cu$_6$Sn$_5$ phase) is present in the Cu–Sn system [3-5]. However, an important question remains unanswered with this consideration. Diffusion coefficients measured in an incremental diffusion couple of Cu/Cu$_6$Sn$_5$ in which only the Cu$_3$Sn phase grows at the interface indicates that Cu has much higher (almost ~30 times) diffusion rate compared to Sn in the Cu$_3$Sn phase [3]. It means that the growth rate of Cu$_3$Sn phase from the Cu$_3$Sn/Cu$_6$Sn$_5$ (interface II) must be much higher compared to the small growth rate of this phase from the Cu/Cu$_3$Sn (interface I). The Kirkendall marker plane (K$_1$ in Figure 2a) could be absent in this phase only if the consumption rate of Cu$_3$Sn at the interface II is very high (*i.e.* $\Delta x_{Cu_3Sn}^{II}$ has a negative value). It would mean that Cu$_6$Sn$_5$ consumes the whole amount of this sublayer of Cu$_3$Sn that is grown from the interface II ($\Delta x_{Cu_3Sn}^{II}$) along with some part of the phase which grows from the interface I ($\Delta x_{Cu_3Sn}^{I}$). In such a situation, the overall thickness of the Cu$_3$Sn phase (*i.e.* $\Delta x_{Cu_3Sn} = \Delta x_{Cu_3Sn}^{I} + \Delta x_{Cu_3Sn}^{II}$) should be much less than what is found in the Cu–Sn system. Moreover, the calculation of the thickness of the sublayers utilizing the diffusion coefficients estimated from the incremental diffusion couples indicates that all the four sublayer values ($\Delta x_{Cu_3Sn}^{I}$, $\Delta x_{Cu_3Sn}^{II}$, $\Delta x_{Cu_6Sn_5}^{II}$, $\Delta x_{Cu_6Sn_5}^{III}$) are positive [3]. That means, we should find a bifurcation of the Kirkendall marker plane, one each in the Cu$_3$Sn and Cu$_6$Sn$_5$ phases. Since the growth rate of the Cu$_3$Sn phase is expected to be small from the interface I ($\Delta x_{Cu_3Sn}^{I}$) and the Cu$_3$Sn grains could not be resolved near the Cu/Cu$_3$Sn phase in a SEM micrograph (Figure 1), we extended our investigation to TEM with the expectation that the location of the Kirkendall marker plane might be detected in the Cu$_3$Sn phase based on the presence of a duplex morphology. Dual column FIB (Focused ion beam) starting from 30kV with final thinning at 5kV is used for the sample preparation for transmission electron microscopy. TEM (transmission electron microscope) operating at 300 kV beam energy is employed for acquiring selected–area electron diffraction pattern (DP) along with the corresponding TEM micrographs viz. dark–field (DF) and bright–field (BF) image. Recorded DPs are indexed using JEMS software.

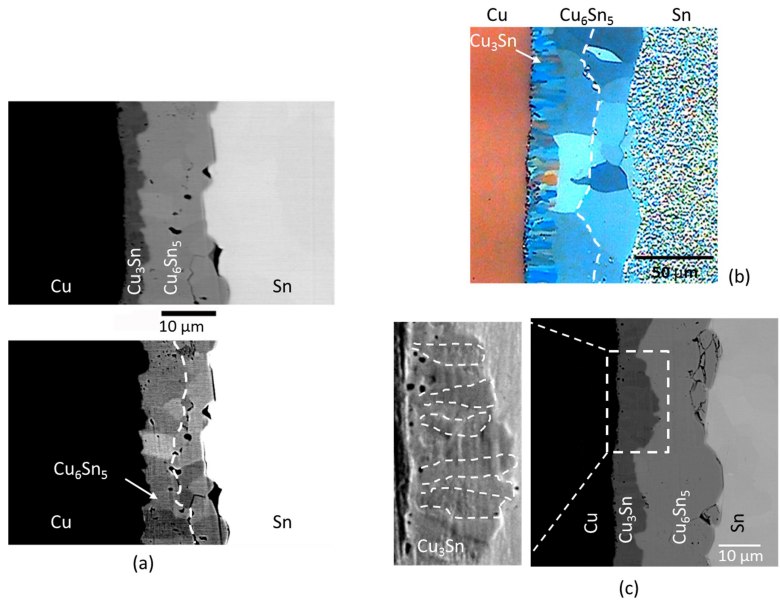

Figure 1: The growth of $Cu_3Sn$ and $Cu_6Sn_5$ phases in the interdiffusion zone of Cu/Sn diffusion couple annealed at (a) 200 °C for 81 hrs (BSE micrograph) (b) 215 °C for 1600 hrs (polarized light optical micrograph) [Courtesy of Dr. A.A. Kodentsov, Eindhoven University of Technology, The Netherlands] (c) 200 °C for 400 hrs, BSE image (right) and focused SE image (left) highlighting elongated $Cu_3Sn$ grains. The location of the Kirkendall marker plane is indicated by duplex morphology inside the $Cu_6Sn_5$ phase as denoted by a dashed line in (a) and (b).

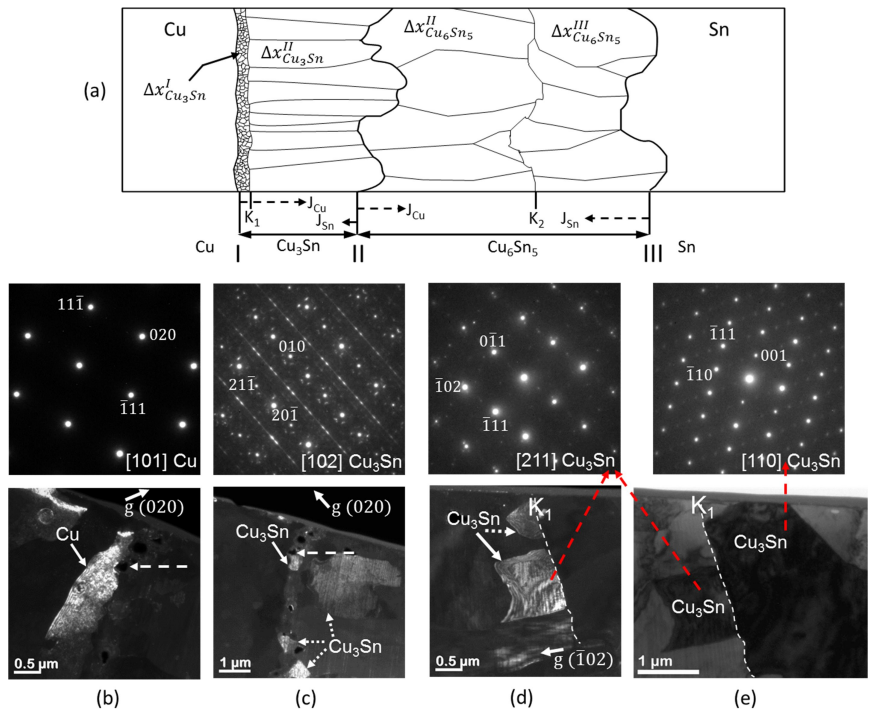

Figure 2: (a) Schematic representation of the growth mechanism of the phases and expected microstructural evolution depending on the location of the Kirkendall marker planes ($K_1$ and $K_2$). Diffraction patterns along with the corresponding TEM micrographs of (b) Cu and (c, d, e) at different locations of the $Cu_3Sn$ phase.

Figure 2 shows microstructures at different locations in the $Cu_3Sn$ phase. Respective DF or BF TEM micrographs along with their indexed DP (recognizing the phases) are shown. The $Cu_3Sn$ phase is an orthorhombic long period superstructure with 80 atoms in a unit cell (oC80, *Cmcm*). It can be seen as combination of 10 units with a prototype of orthorhombic $Cu_3Ti$ (oP8, *Pmnm*) ordered lattice. All the DP of the $Cu_3Sn$ phase is indexed based on oP8 crystal structure [17]. Weak superlattice reflections can be seen in DP of the $Cu_3Sn$ phase due to ordering. Figure 2b shows the DP acquired from one of the Cu grain (near the $Cu/Cu_3Sn$ interface) along with the corresponding DF image. DP can be indexed with zone axis [101] of Cu and the DF image is acquired using (020) reflection. Dashed arrow indicates the same hole (very close to the $Cu/Cu_3Sn$ interface) in both Figure 2b and 2c. On the right side of the interface, many small grains of $Cu_3Sn$ are found, as shown in Figure 2c. DP acquired from grain just below the dashed arrow is indexed with zone axis [102] of the $Cu_3Sn$ phase. Apart from weak superlattice reflections, extra double diffraction spots can be also seen in this DP (Figure 2c). The other grains of the $Cu_3Sn$ phase (indicated by dotted arrows), which might be oriented along the same zone axis [102], can be seen in the DF image (at lower magnification) acquired using (020) reflection. Therefore, there are many small grains of the $Cu_3Sn$ phase close to the $Cu/Cu_3Sn$ interface and following relatively bigger grains are found. Figure 2d shows the DP acquired from the grain (indicated by the arrow) along with the corresponding DF image. DP is indexed with zone axis [211] of the $Cu_3Sn$ phase. Other grain of the $Cu_3Sn$ phase (indicated by a dotted arrow) might also be oriented along the same zone axis [211] as can be seen in the DF image acquired using ($\bar{1}$02) reflection. These grains (Figure 2d) are relatively bigger than the grains observed close to $Cu/Cu_3Sn$ interface (Figure 2c). However, overall, all these grains are much smaller than the grains detected in the SEM micrograph in Figure 1c. Therefore, if the Kirkendall marker plane ($K_1$) in the $Cu_3Sn$ phase exist, it could be demarcated by two sublayers, one with smaller grains and another with much bigger grains. This is indeed found, as shown in the BF micrograph in Figure 2e. The DF and BF images, as shown in Figure 2d and 2e, are acquired from the same region. DP acquired from the very big grain (indicated by a red dotted arrow in Figure 2e) is indexed with zone axis [110] of the $Cu_3Sn$ phase. This means that the Kirkendall marker plane is indeed present in this phase, which demarcates two different sublayers with different grain morphologies. The sublayer between interface I ($Cu/Cu_3Sn$) and $K_1$ has many small grains indicating the repeated nucleation of grains along with growth of this sublayer because of diffusion of Sn originated from the interface II ($Cu_3Sn/Cu_6Sn_5$). Another sublayer between $K_1$ and interface II has much bigger grains covering the thickness of whole sublayer, which is grown because of reaction (of Cu with $Cu_6Sn_5$) and dissociation (of Sn from $Cu_6Sn_5$) at the interface II. Instead of nucleating repeatedly, the product phase continuously joins with the existing $Cu_3Sn$ grains in this sublayer. This indicates that the nucleation of $Cu_3Sn$ might be easier at the interface I (on Cu) as compared to the interface II (on $Cu_6Sn_5$). Most of the intermetallic compounds in various systems grow with large grains covering the thickness from one interface to the Kirkendall marker plane or one interface to another (if the Kirkendall marker plane is not present or it is present very close to one of the interfaces [18]). However, there are few systems [19, 20], in which grains in a particular sublayer of intermetallic phase grow with many smaller grains indicating the ease of nucleation similar to what we have found between interface I and $K_1$. In a previous study based on the TEM analysis, Tian et al. [21] found that the Kirkendall marker plane (detected by inert marker) in the $Cu_2(In,Sn)$ phase indeed demarcates two sublayers with different types of grain morphologies which support the concept of the physico–chemical approach [2].

Now the question is why the bifurcation of the Kirkendall marker plane in the Cu–Sn system could not be detected when inert particles were used as Kirkendall markers in the previous studies [3-5]. It should be noted here that a system might fulfill the conditions for bifurcation of the marker plane, however, the markers (inert particles) will be able to split into two different phases only if both the phases start growing together at the very initial stage of diffusion annealing. Bulk diffusion couples, in general, show the simultaneous growth, however, the sequential growth of the phases in this system is reported [22], which was indeed suspected based on the thermodynamic viewpoint [3]. The $Cu_3Sn$ phase grows only after the growth of $Cu_6Sn_5$ phase. Therefore, once the markers (inert particles) are trapped inside the $Cu_6Sn_5$ phase, they cannot move in the $Cu_3Sn$ phase. On the other hand, the microstructural analysis has evolved as a reliable technique for the detection of the Kirkendall marker plane [1]. This is now commonly practiced in the systems in which markers cannot be used such as material in applications or thin films [1]. This is also recently followed in many refractory metal–silicon systems [18] since the diffusion couples could not be bonded successfully when inert marker particles were used at the interface between these hard materials, often creating a gap between them.

Now we estimate the diffusion parameters, to facilitate our discussion on the growth of Kirkendall voids, in the next section. The integrated interdiffusion coefficients of the phases with narrow homogeneity range in this system are estimated following the relation developed by Wagner [23]

$$\widetilde{D}^{\beta}_{int} = \frac{(N_B^{\beta} - N_B^-)(N_B^+ - N_B^{\beta})}{(N_B^+ - N_B^-)} \frac{(\Delta x^{\beta})^2}{2t} + \frac{\Delta x^{\beta}}{2t} \left[ \left(\frac{N_B^+ - N_B^{\beta}}{N_B^+ - N_B^-}\right) \int_{x^{-\infty}}^{x^{\beta_1}} \frac{V_m^{\beta}}{V_m}(N_B - N_B^-)dx + \left(\frac{N_B^{\beta} - N_B^-}{N_B^+ - N_B^-}\right) \int_{x^{\beta_2}}^{x^{+\infty}} \frac{V_m^{\beta}}{V_m}(N_B^+ - N_B)\,dx \right]$$

(1)

where β is the phase of interest, $N_B^-$ and $N_B^+$ are the compositions of left– and right–hand side of end–members *i.e.* unaffected parts of the diffusion couple located at $x^{-\infty}$ and $x^{+\infty}$, respectively. $\Delta x^{\beta}$ is the thickness, $N_B^{\beta}$ is the average or stoichiometric composition of the phase of interest, $V_m$ is the molar volume, and *t* is the annealing time. The steps for estimation of these data can be found in the text book given in Ref. [1]. The molar volume of the phases are $V_m^{Cu_3Sn} = 8.59 \times 10^{-6}$ m$^3$/mol and $V_m^{Cu_6Sn_5} = 10.59 \times 10^{-6}$ m$^3$/mol [3]. For our calculations, we considered different types (with different annealing times) and different locations of the diffusion couple. The average values are estimated as (which are similar to the values estimated earlier [3])

$\widetilde{D}^{Cu_3Sn}_{int} = (1.2 \pm 0.2) \times 10^{-17}\ m^2/s$

$\widetilde{D}^{Cu_6Sn_5}_{int} = (8.5 \pm 1) \times 10^{-17}\ m^2/s$

The ratio of the tracer diffusivities, $D_i^*$ are estimated following van Loo's method [24]

$$\frac{D_B^*}{D_A^*} = \left[ \frac{N_B^+ \int_{x=-\infty}^{x_K} \frac{Y_B}{V_m} dx - N_B^- \int_{x_K}^{x=+\infty} \frac{(1-Y_B)}{V_m} dx}{-N_A^+ \int_{x=-\infty}^{x_K} \frac{Y_B}{V_m} dx + \bar{A} \int_{x_K}^{x=+\infty} \frac{(1-Y_B)}{V_m} dx} \right] \quad (2)$$

where $Y_B = \frac{N_B - N_B^-}{N_B^+ - N_B^-}$ and $x_K$ is the location of the Kirkendall marker plane. Note here that we have neglected the role of vacancy–wind effect [1] since the structure factor required for the estimation of this effect is not known. Further, we estimate the ratio of tracer diffusivities, since the partial molar volumes of the components ($\bar{V}_i$) in these phases are not known. Again considering different types of diffusion couples we estimated the values as

$$\left[\frac{D_{Sn}^*}{D_{Cu}^*}\right]_{K_1(Cu_3Sn)} = 0.033 \quad i.e. \quad \left[\frac{D_{Cu}^*}{D_{Sn}^*}\right]_{K_1(Cu_3Sn)} = 30 \pm 10$$

$$\left[\frac{D_{Sn}^*}{D_{Cu}^*}\right]_{K_2(Cu_6Sn_5)} = 2.3 \pm 1$$

It should be noted here that the calculation of tracer (or intrinsic) diffusion coefficients by diffusion couple technique introduces high error when the ratio is outside the range of 0.1–1 [1]. Moreover, the error in calculation is high especially in this system because of the waviness of phase layers. When a single phase layer grows in an incremental diffusion couple, $Cu_3Sn$ grows with more or less flat interfaces, however, both the interfaces of $Cu_6Sn_5$ are found to be highly wavy [3]. Therefore, $Cu_3Sn$ is flat at the interface I, however, very wavy at the interface II, since the sublayer that is grown from interface II is dependent on the growth of $Cu_6Sn_5$, as discussed before. Thermodynamics constrain the system for the growth with flat interfaces [24]; however, orientation–dependent anisotropic growth [24] of $Cu_6Sn_5$ makes it wavy which must be clear from grains revealed in this phase, as shown in Figure 1. Nevertheless, the experimental evidence of finding the bifurcation of the Kirkendall marker plane in the Cu–Sn system will bring finer understanding on the growth of the phases based on the theoretical analysis on vacancy creation and annihilation in this technologically important system, which draws special attention in many groups [25-28].

## 2.2   The effect of impurity concentration in Cu on the growth of Kirkendall voids in the $Cu_3Sn$ phase

From the estimated ratio of tracer diffusion coefficients $\left[\frac{D_{Cu}^*}{D_{Sn}^*}\right]$, it is clear that Cu has much higher diffusion rate compared to Sn in the $Cu_3Sn$ phase, whereas, this difference is not that high in the $Cu_6Sn_5$ phase. Considering a constant molar volume of a phase, the flux of vacancies ($J_V$) can be related to the intrinsic fluxes ($J_i$) of components by $J_V = -(J_{Cu} + J_{Sn})$. Note here that $J_{Cu}$ and $J_{Sn}$ have opposite signs (Figure 2a). Therefore, the flux of vacancies is significantly higher in the $Cu_3Sn$ phase. In fact, even 1% relative excess non–equilibrium vacancy concentration can be enough to create voids if not absorbed by the sinks [11]. It is also apparent that the vacancies are not absorbed completely in the $Cu_3Sn$ phase since Kirkendall voids (could be recognized as dark spots) are found in this phase, as shown in Figure 3. It is very common to find cracks in the brittle $Cu_6Sn_5$ phase, which are introduced during cross–sectioning of the diffusion couple and metallographic preparation. Since the main focus in this section is to discuss the growth of voids in the $Cu_3Sn$ phase, henceforth the micrograph of this phase is only shown, the $Cu_6Sn_5$ phase is shown only when it is required for any other discussion. The microstructures of the interdiffusion zone are examined using scanning electron microscope equipped with field emission gun (FE–SEM) in both BSE (back–scattered electron) and SE (secondary electron) imaging mode.

The presence of impurities plays a very important role in the growth of Kirkendall voids. For example, it is known that the presence of impurities can increase the concentration of vacancies by decreasing the enthalpy of vacancy formation [29]. Because of relevance to the manufacturing process of making contacts between Cu as one of the layers of UBM and Sn–based solder, most of the studies [6-8] on the growth of Kirkendall voids in the $Cu_3Sn$ phase are mainly focused on electroplated (EP) Cu. Generally, for a smooth and bright layer, different additional constituents (both organic and inorganic) are used in Cu electroplating bath. which are known to promote the inclusion of fairly high concentration of impurities in Cu such as S, Cl, C, O, N and H. Out of these, S if found to play an adverse role in the growth of Kirkendall voids [6, 30]. Interestingly, in few studies, it is reported that the Kirkendall voids are not found or found with very small numbers when Cu with higher purities are coupled with Sn or Sn–based solder [31, 32]. It is indeed expected to find the low growth of the voids because of the smaller concentration of impurities in Cu. However, according to the certificate provided by Alfa Aesar (USA), the S content is around 0.2 ppm in 99.9999 wt.% Cu, which is used in this study. Therefore, the voids are not expected to be completely absent even when the very high purity of Cu is used. In this study, we have considered different purities of Cu starting from 99.9999 wt.% to commercially pure for the comparison of the growth of voids with EP Cu.

As already explained in the previous section, the $Cu_3Sn$ phase grows by reaction–dissociation of components from the $Cu/Cu_3Sn$ and $Cu_3Sn/Cu_6Sn_5$ interfaces, whereas, the $Cu_6Sn_5$ phase grows from the $Cu_3Sn/Cu_6Sn_5$ and $Cu_6Sn_5/Sn$ interfaces (Figure 2a). Therefore, it is expected that the condition of Sn should not affect the growth of Kirkendall voids in the $Cu_3Sn$ phase directly or significantly. To cross–check this statement, one diffusion couple is prepared between bulk 99.9 wt.% Cu and EP Sn (Figure 3b) for comparison of the growth of the $Cu_3Sn$ phase with a diffusion couple of bulk 99.9 wt.% Cu and bulk 99.99 wt.% Sn (Figure 3a). After comparison of many voids at different locations in the interdiffusion zone, we did not find any significant differences. Therefore, now onwards, the results for the different purity of Cu and bulk 99.99 wt.% Sn are shown. Electroplating of Sn and Cu is conducted in an air–conditioned (AC) room maintained at $20 \pm 5$ °C with a current density of 20 mA/cm$^2$. Sn electroplating solution used in this study contains $SnSO_4$, $H_2SO_4$, and Stannolume additive and brightener [33]. Following, the practice in electronic industries, the commercially available Cu electroplating bath provided by Grauer & Weil (Growel, India) is used in this study, which contains $CuSO_4$, $H_2SO_4$, HCl, and different Cuprobrite additive and brightener [34]. Since, the impurity concentration in EP Cu is found to be much higher than 99.9 wt.%, commercially pure Cu is considered for the sake of comparison of bulk Cu and EP Cu.

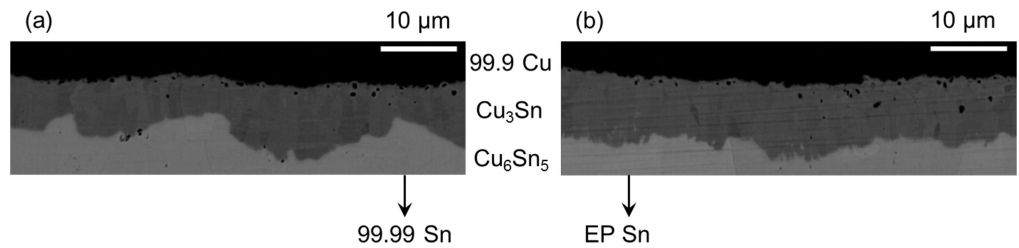

Figure 3: BSE micrographs with comparison of the growth of the Kirkendall voids in diffusion couples of (a) 99.9 wt.% bulk Cu and 99.99 wt.% bulk Sn, and (b) 99.9 wt.% bulk Cu and electroplated (EP) Sn annealed at 200 °C for 400 hrs.

When the different purities of Cu such as 99.9999, 99.999 and 99.9 wt.% are coupled with 99.99 wt.% Sn, the thicknesses of the phase layers are found to be more or less the same for a particular annealing time indicating not much difference in the growth rate of both the phases. However, there is a difference in the growth of Kirkendall voids in the $Cu_3Sn$ phase. Figure 4a shows the voids for different purities of Cu after annealing for 144 hrs at 200 °C. The $Cu_3Sn$ phase is more or less free from voids for 99.9999 wt.% Cu. A small number of voids are found when the diffusion couple is prepared with 99.999 wt.% Cu. However, they are not distributed evenly. Voids are found only at a few places, which indicates that impurities might not be distributed evenly. With the increase in impurity of Cu to 99.9 wt.%, the number of voids are found to be much higher and also distributed more or less evenly. Hence, to facilitate our systematic analysis with higher annealing time experiments, we have chosen only extremes cases of Cu purities *i.e.* 99.9999 and 99.9 wt.%. After increasing the annealing time to 400 hrs (at the same temperature), as shown in Figure 4b, the Kirkendall voids are found for both the purities of Cu although the number of voids is significantly higher for 99.9 wt.% Cu.

To get further insights on the growth of Kirkendall voids, the void statistics *i.e.* distribution of number of voids and their sizes viz. equivalent diameter (in terms of average size with ± 0.05 μm range) are estimated by using MIPAR (Materials Image Processing and Automated Reconstruction) [35], a powerful image analysis software [36]. An example of detection of these voids by the software (for 99.9 wt.% Cu diffusion couple annealed for 400 hrs at 200 °C, *i.e.*, Figure 4b) is shown in Figure 4c. First, the $Cu_3Sn$ phase boundary can be identified in the MIPAR software as shown by green color line and then the voids inside the phase can be detected as shown by white color lines. To avoid error, we have not considered the void sizes less than 0.15 μm. Moreover, all the figures are checked carefully for the voids to be selected correctly, by making use of some of the inherently inbuilt simple and unique features in the MIPAR software. After the analysis, if any wrong consideration of other black spots as voids is identified, the software allows to remove them manually from the list of equivalent diameters estimated. In fact, it is also possible to view the both viz. original image (Figure 4b) and image with detected voids (Figure 4c) together at the same time to pinpoint if the voids are detected correctly. Measurements are done using many images such that at least 200 voids are considered for each condition and then these are normalized for 1000 $\mu m^2$ area of the $Cu_3Sn$ phase.

To understand the effect of annealing times (for the same Cu purity) on the growth of Kirkendall voids, the void size and number distributions are compared for 99.9 wt.% Cu after annealing for 144 and 400 hrs (at the same temperature), as shown in Figure 5a. The total number of voids are found to be 136 and 150 per 1000 $\mu m^2$ after annealing for 144 and 400 hrs respectively. The maximum size of a void is found to be around 0.9 µm after 144 hrs and 1.1 µm after 400 hrs for 99.9 wt.% Cu. Moreover, the number of voids of bigger sizes increase with the increase in annealing time. Overall number and size of voids will depend on the (i) relaxation of vacancies by sinks (K sinks) [37] (ii) heterogeneous nucleation of new voids because of supersaturation of vacancies and (iii) relaxation of vacancies in existing voids (F sinks) [37]. Relaxation of vacancies by sinks will not create a void. Nucleation of new voids will increase the number of voids, whereas, the addition of vacancies in the existing void will increase the overall size of a void. Therefore, we can conclude that the excess vacancies, which are not absorbed by sinks, can generate new voids as well as increase the size of existing voids with the increase in annealing time.

To understand the effect of impurity in Cu (after the same annealing time) on the growth of Kirkendall voids, voids distributions are plotted in Figure 5b for 99.9999 and 99.9 wt.% Cu. Both the diffusion couples are annealed for the same time of 400 hrs. The clear differences in void size distribution can be seen in the micrographs of the $Cu_3Sn$ phase, as shown in Figures 4b. Further insights on the same (*i.e.* growth of voids) are given by the distribution plot. The total number of voids for 99.9999 wt.% Cu is found to be 88 per 1000 $\mu m^2$ compared to 150 voids for 99.9 wt.% Cu after annealing for 400 hrs. The estimated number of voids of smallest size range (*i.e.* the average equivalent diameter of 0.2 µm) for 99.9999 wt.% Cu is found to be very less, indicating the lower rate of nucleation of voids which might be due to lack of availability of impurities for nucleation. Therefore, excess vacancies might prefer to join the existing voids. Void size and number distribution for 99.9 wt.% Cu indicates that voids could nucleate with a higher rate in this case and even the flux of excess vacancies must be much higher such that numbers of almost all sizes of voids are higher for 99.9 wt.% Cu when compared to the voids for 99.9999 wt.% Cu.

Next, we consider the growth of Kirkendall voids for electroplated (EP) Cu, as shown in Figure 6a. Since the growth rate of the Kirkendall voids is much higher in this case, the annealing time is restricted to 100 hrs at 200 °C. As shown in Figure 5c, void size distribution is compared with the highest impurity of bulk Cu considered till now *i.e.* 99.9 wt.% Cu, which was annealed for 144 hrs (at the same temperature). Although the diffusion couple prepared with EP Cu is annealed for a smaller time (Figure 6a) compared to 99.9 wt.% Cu (Figure 4b), the void numbers and sizes are clearly much higher for EP Cu which can also be understood from the void distribution plot. The total number of voids is estimated as 435 per 1000 $\mu m^2$ for EP Cu after 100 hrs of annealing compared to 136 voids for 99.9 wt.% Cu after 144 hrs of annealing. The highest average void size (equivalent diameter) is found to be 1.3 µm for EP Cu. For a reasonable comparison, considering different annealing times of these diffusion couples in which the growth of phases follow a parabolical relation with time [4], the void numbers are plotted with respect to the equivalent diameter ($d$) normalized by the square root of the annealing time *i.e.* $d/t^{1/2}$ (μm/hr$^{1/2}$) as shown in Figure 5d. This brings even higher difference for these two different impurities of Cu. Therefore, we can safely state that the impurity concentration in EP Cu must be much higher than 99.9 wt.% Cu. Shimizu et al. [38] stated based on their analysis that the total concentration of impurities in EP Cu could be close to 1 wt.%, which means the purity of EP Cu could be much higher than 99.0 wt.% Cu.

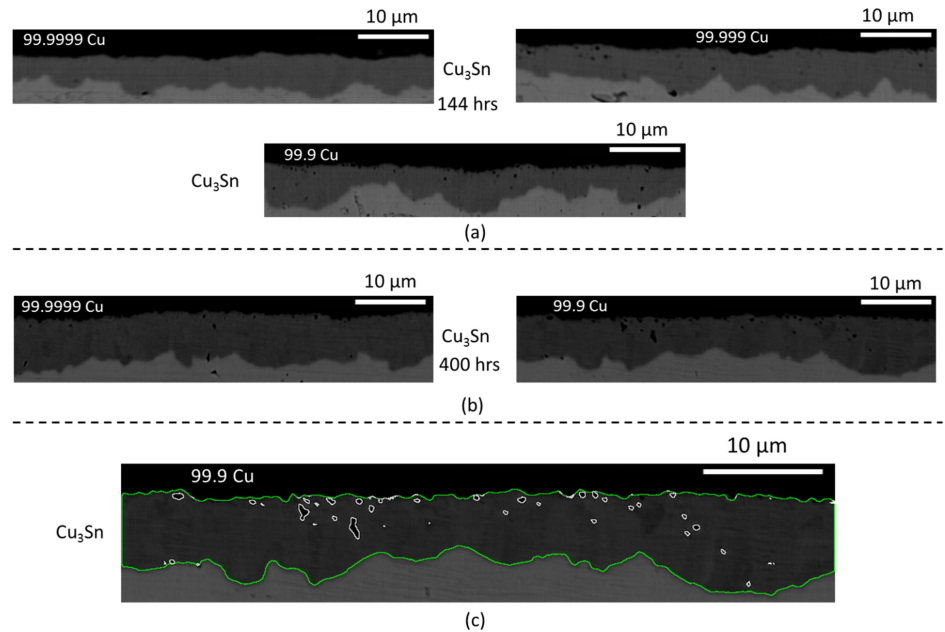

Figure 4: BSE micrograph of the Cu$_3$Sn phase in the interdiffusion zone of Cu/Sn bulk diffusion couples annealed at 200 °C. Diffusion couples of (a) 99.9999, 99.999 and 99.9 wt.% Cu with 99.99 wt.% Sn annealed for 144 hrs, and (b) 99.9999 and 99.9 wt.% Cu with 99.99 wt.% Sn annealed for 400 hrs. (c) An example of analysis of the Kirkendall void size distribution by MIPAR on the micrograph of 99.9 wt.% Cu shown in Figure (b).

Considering some of the embedded impurities in EP Cu could be volatile in nature (which may be released during annealing under vacuum), the effect of vacuum pre–heat treatment of the Cu on the growth of the Kirkendall voids in the Cu$_3$Sn phase is studied previously [7, 39]. This could be beneficial if the temperature of heat treatment is restricted to the limit such that an electronic component can withstand it. Yin and Borgesen [7] heat treated EP Cu at 650 °C and found the negligible growth of the Kirkendall voids after that. Similar experimental observations are reported at 400–600 °C by Kim and Yu [39]. This indicates that Cu is indeed purified with a pre–heat treatment step. However, the temperature of the annealing is very high in both these studies [7, 39]. In the present study, to examine if this beneficial step is useful even at lower pre–heat treatment temperature, we first heat treated EP Cu foil in high vacuum (~$10^{-4}$ Pa) at the same temperature and time as that of diffusion annealing *i.e.* at 200 °C for 100 hrs. This is designated as EP (HT) in this manuscript. Following, this is electroplated with Sn to prepare a diffusion couple (similar to a couple of EP Cu and EP Sn shown in Figure 6a). This is further annealed at 200 °C 100 hrs for studying a difference in the growth of Kirkendall voids in these both $Cu_3Sn$ phase layers. It can be seen in Figure 6b that the voids are mostly accumulated near the $Cu/Cu_3Sn$ interface for EP (HT) Cu. This is much clearer in the SE image, as shown in Figure 6c. A previous study by Singh et. al. [40] indicates that impurities are transported to the Cu surface (leaving Cu beneath the surface as pure) following heat treatment of Cu alone and this could be the reason to find all the voids at the $Cu/Cu_3Sn$ interface. Although it is impossible to estimate voids distribution for EP (HT) Cu, however looking at the microstructure (Figure 6b and 6c), it seems that the number of voids could be less for EP (HT) Cu when compared to EP Cu as plated (Figure 6a). This indicates the decrease in impurity concentration because of vacuum pre–heat treatment step. This further indicates the significant role of impurities on the growth of Kirkendall voids, since in the case of EP Cu (Figure 6a) voids are spread over the whole $Cu_3Sn$ phase layer while in the case of EP (HT) Cu (Figure 6b and 6c) voids are negligible inside the $Cu_3Sn$ phase (and found only at $Cu/Cu_3Sn$ interface). The accumulation of voids for EP (HT) Cu makes the interface very weak, which is not good for the electro–mechanical reliability of an electronic component during service.

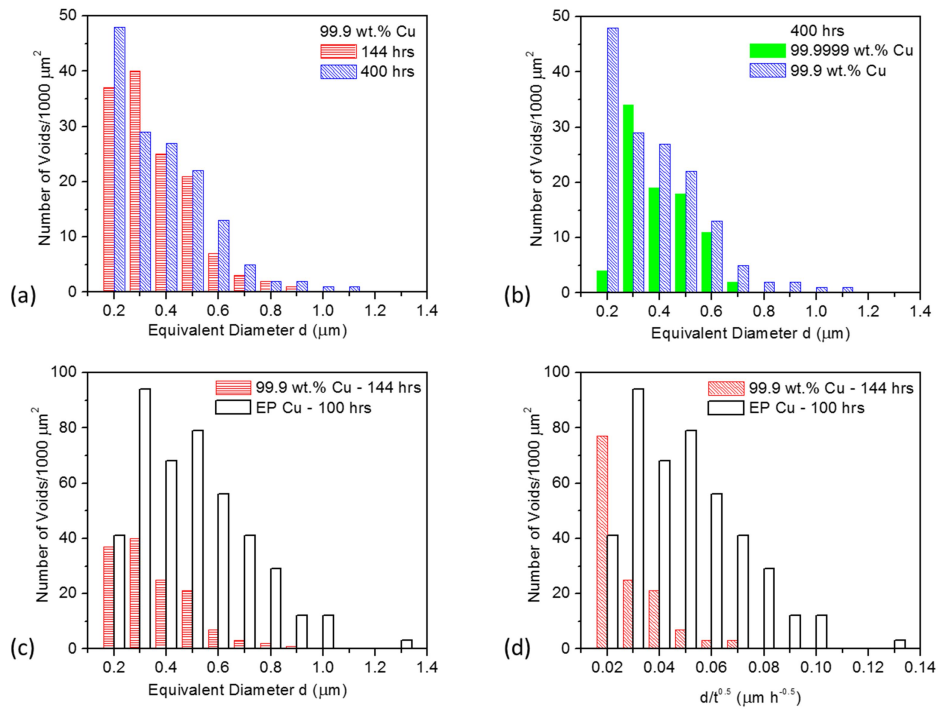

Figure 5: Void size distribution (at 200 °C) comparing (a) two annealing times, 144 and 400 hrs for 99.9 wt.% Cu (b) two different purities, 99.9999 and 99.9 wt.% Cu for 400 hrs (c) two different purities viz. 99.9 wt.% Cu and EP Cu for two different annealing times, and (d) 99.9 wt.% Cu and EP Cu with respect to the normalized time.

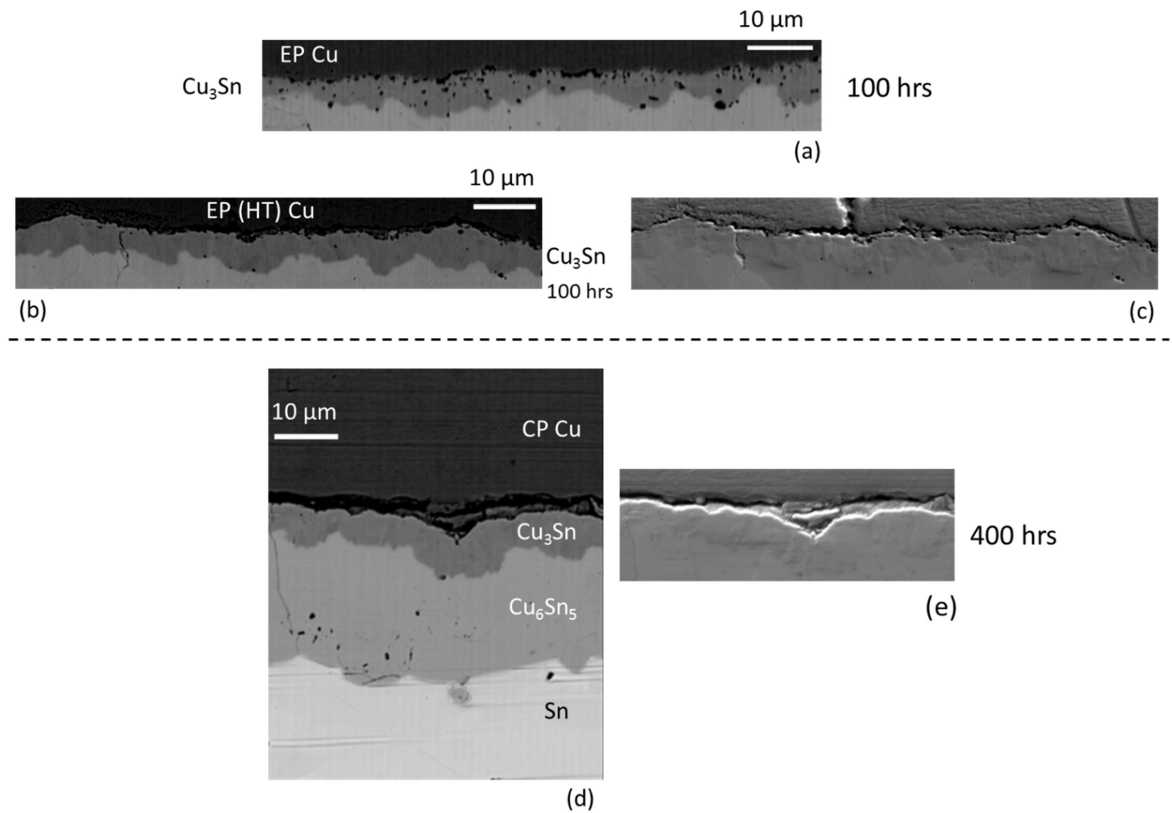

Figure 6: Micrographs of the interdiffusion zone of Cu/Sn diffusion couple annealed at 200 °C for 100 hrs (a) EP Cu with EP Sn, BSE image, (b) EP (HT) Cu with EP Sn, BSE image (c) SE image of the same. HT refers to vacuum pre–heat treatment step. (d) BSE image showing both the $Cu_3Sn$ and $Cu_6Sn_5$ phases in a diffusion couple of Commercial Pure (CP) Cu with 99.99 wt.% bulk Sn annealed at 200 ºC for 400 hrs and (e) corresponding SE image showing only the $Cu_3Sn$ phase.

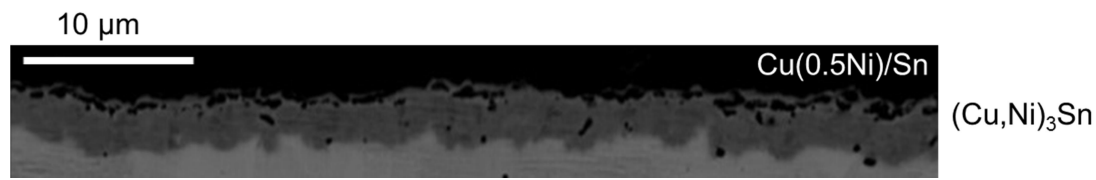

Figure 7: BSE micrograph for the comparison of growth of Kirkendall voids in diffusion couple of Cu(0.5Ni)/Sn annealed at 200 °C for 81 hrs.

To compare these results with bulk Cu, since the void concentration for EP Cu is found to be higher than 99.9 wt.% bulk Cu, commercial pure (CP) Cu which is known to have the impurity content of around 1–2 wt.% [41, 42] is coupled with bulk Sn and annealed at 200 °C for 400 hrs, as shown in Figure 6d and 6e. Looking at the back–scattered electron (BSE) image (Figure 6d), it feels as if there is a gap at the $Cu/Cu_3Sn$ interface. However, the secondary electron (SE) image, as shown in Figure 6e, clearly shows that there is actually a thin crack or fine separation present at the $Cu/Cu_3Sn$ interface. The Kirkendall voids are not found inside the $Cu_3Sn$ phase, which indicates that all the voids are accumulated at the interface for CP Cu creating bigger holes and making the interface very weak. This might cause the separation of Cu/Cu$_3$Sn interface during cross–sectioning of the diffusion couple and metallographic preparation. Considering the thickness of the Cu$_3$Sn phase after including the black contrast in the BSE image (Figure 6d) for CP Cu (since SE image shown in Figure 6e indicates the presence of the Cu$_3$Sn phase almost up to the Cu/Cu$_3$Sn interface), we find the thicknesses of both the phase layers as more or less the same for CP Cu when compared to 99.9 wt.% Cu (Figure 1c) after annealing for 400 hrs (at the same temperature). It indicates that the presence of the Kirkendall voids at the interface did not affect the supply of Cu from the end–member. It is well possible that the decrease in cross–sectional area is compensated by the surface diffusion of Cu through the voids.

We conducted another experiment with the addition of 0.5 at.% Ni (99.95 wt.% pure) in 99.999 wt.% Cu, as shown in Figure 7. The Cu(0.5Ni) alloy was prepared in an arc melting unit under an argon atmosphere and subsequently homogenized at 1050 °C for 50 hrs. The growth of voids in Ni–free 99.999 wt.% Cu can be seen in Figure 4a. The addition of Ni in higher amount is avoided since it affects the growth kinetics of the Cu$_3$Sn phase [43, 44]. It should be noted here that Ni is used as another UBM along with Cu and, therefore, it is important to study the effect of Ni addition on the growth of voids. The growth of voids increases drastically with Ni addition. It is generally believed that S and other (organic and inorganic) impurities, which are added during electroplating of Cu, play an important role in the significant increase in growth rate of the Kirkendall voids. It is also speculated that voids created in such a situation are not grown because of the Kirkendall effect but because of the presence of organic impurities (or its complex) in electroplated Cu [7]. In this study, we find that the addition of 0.5 at.% Ni in bulk Cu, which is equivalent to the concentration of impurities incorporated in an electroplated Cu, produces the voids with comparable rates. Therefore, the growth of these voids must be influenced by the presence of other inorganic impurities also along with the organic impurities. This further indicates that the different impurities and alloying addition can play an adverse role in the growth of Kirkendall voids.

## 3. Conclusions

For the very first time, a systematic study is conducted on the two very important aspects related to the Kirkendall effect in a technologically important Cu–Sn system: (a) to detect bifurcation of the Kirkendall marker plane based on the microstructural evolution (b) to study the role of impurities on the growth rate of Kirkendall voids in the Cu$_3$Sn phase. Much finer understanding on the growth process of the phases based on microstructural evolution and Kirkendall voids are developed based on this study. Few keys findings are summarized as:

(i) The experimental evidence of the presence of the bifurcation of the Kirkendall marker plane is shown in the Cu–Sn system. The splitting of the markers was predicted before by Paul et al. [3]; however, the conventional marker experiment utilizing inert particles failed to detect this [3-5]. In this study, the locations of these planes in both the phases viz. Cu$_3$Sn and Cu$_6$Sn$_5$ are detected by the microstructural evolution. The reason for not detecting this by the conventional marker experiment is explained.

(ii) This study strengthens the concept of the physico–chemical approach [2], which explains that one can efficiently detect the locations of the Kirkendall marker planes based on the analysis of microstructural evolution. There is no need to use inert particles at the interface. This is important since one cannot easily introduce inert particles in a system prepared with thin films or the materials in applications. The growth mechanism of the phase based on the relative mobilities of components can still be understood by locating this plane based on the microstructural evolution, which was not possible earlier.

(iii) Quantitative diffusion analysis indicates that there is not much difference in the diffusion rates of the components in the $Cu_6Sn_5$ phase. On the other hand, Cu has almost 30 times higher diffusion rate compared to Sn in the $Cu_3Sn$ phase. Since the Kirkendall voids grow in the $Cu_3Sn$ phase, it is evident that the flux of vacancies created due to the difference in diffusion rates of components are not absorbed completely by the sinks.

(iv) The sublayer between the $Cu/Cu_3Sn$ interface and $K_1$ in the $Cu_3Sn$ phase grows by reaction of diffusing Sn with Cu at this interface (interface I). Sn is generated by dissociation of $Cu_6Sn_5$ at the $Cu_3Sn/Cu_6Sn_5$ interface leading to the growth of $Cu_3Sn$ in another sublayer between $K_1$ and this interface (interface II). This sublayer is also grown because of diffusion of Cu from the interface I at much higher rate and then by reaction with $Cu_6Sn_5$ at the interface II. As a result, the thickness of the sublayer that is grown from Cu at the interface I is much smaller compared to the thickness of another sublayer that is grown from $Cu_6Sn_5$ at the interface II.

(v) Consequently, as found in the present study, the growth of the voids in the sublayer that is grown from Cu is higher because of the presence of impurities in Cu and high flux of excess vacancies near the interface I, which are not absorbed by the sinks. Since the voids could grow even in the other sublayer that is grown from $Cu_6Sn_5$ at the interface II, it is evident that impurities could transport inside the product phase from the Cu end–member and the concentration of excess vacancies must be enough in this sublayer also which is required for the formation and growth of the voids. This is important for the theoretical studies, since the growth of voids in the sublayer that is grown from $Cu_6Sn_5$ is not considered or explained in the theoretical studies published earlier by others.

(vi) The presence of impurities plays an important role in the growth of voids. The growth rate increases significantly when the impurity concentration is equal to or more than 0.1 wt.% in Cu. The presence of a very high number of voids for EP Cu indicates the incorporation of fairly high concentration of impurities during electroplating. Voids size distribution indicates the nucleation of new voids can happen along with the growth of the existing voids.

(vii) A theoretical study by Svoboda and Fisher [25] indicates that an incoherent interface acts as non–ideal source and sink for vacancies leading to most effective site for nucleation of voids. These are detached from the interface during interface migration and stay dispersed inside the product phase. Surface diffusion because of the presence of pores is also possible such that voids are detached from the

(viii) interface after growth of fresh $Cu_3Sn$ at the $Cu/Cu_3Sn$ interface, which is recently shown by Gusak et al. [45].

(viii) The role of surface diffusion of both the components is evident from the fact that the growth rates of the phases are not much different even when many voids are accumulated mainly near to the $Cu/Cu_3Sn$ interface in the case of CP Cu leading to decrease in the interfacial area through which the components diffuse. However, at this point it is not clear to us that why the voids spread over the whole phase layer for 99.9999–99.9 wt.% Cu and EP Cu and these are accumulated near the interface for EP(HT) Cu and CP Cu.

This system draws a special attention for theoretical analysis [25, 37, 45, 46] because of technological importance. The finding of the bifurcation of the Kirkendall marker plane and the systematic analysis of the growth of the Kirkendall voids with increasing concentration of impurity will help to establish the exact underlying mechanism based on simulations, which is otherwise difficult to establish based on just experimental studies.

**Acknowledgments**

Authors would like to acknowledge the help of Dr. John M. Sosa, The Ohio State University, Columbus, USA and his team for their kind help and support for the use of MIPAR in the present work. We also acknowledge the support of staff members in MNCF (Micro Nano Characterization Facility) and AFMM (Advanced Facility for Microscopy and Microanalysis) facilities at Indian Institute of Science (IISc) Bangalore for providing the usage of FIB and TEM respectively for this study.